\documentclass[10pt]{article}
\usepackage{times}
\usepackage{a4wide}
\usepackage{amsfonts}
\usepackage{amssymb}
\usepackage{multicol}
\usepackage{amsmath}
\usepackage{graphicx}
\usepackage[usenames,dvipsnames]{xcolor}
\usepackage{ifpdf}
\ifpdf
\usepackage[pdftex,unicode,implicit]{hyperref}

\usepackage[condensed,math]{anttor}
\usepackage[T1]{fontenc}

\DeclareSymbolFont{extraup}{U}{zavm}{m}{n}
\DeclareMathSymbol{\varheart}{\mathalpha}{extraup}{86}
\DeclareMathSymbol{\vardiamond}{\mathalpha}{extraup}{87}

\usepackage[sort&compress,comma]{natbib}
\bibpunct{[}{]}{,}{n}{}{,}

\hypersetup{%
  pdftitle    = { },
  pdfkeywords = {string theory, black holes, Calabi Yau, compactification, quantum corrections, supergravity model, Lambert function, no-hair theorem},
  pdfauthor   = {Pablo Bueno and C. S. Shahbazi},
  plainpages  = true,
  colorlinks  = true,
  citecolor   = Red	,
  urlcolor    = Green,
  linkcolor   = Blue	
}

\else
  \usepackage[dvips]{graphicx}
  \usepackage[unicode,implicit]{hyperref}

\fi

\makeatletter
\@addtoreset{equation}{section}
\makeatother

\begin{document}
\vspace{1cm}
\hfill \hspace{-0.3cm} IFT-UAM/CSIC-13-107

\begin{center}
\LARGE{\vspace{2cm }\bf\hspace{-0.1cm} The violation of the No-Hair Conjecture in four-dimensional ungauged Supergravity\\}
\vspace{1cm}
\renewcommand{\thefootnote}{\alph{footnote}}
{\large \textcolor{Black}{Pablo~Bueno}$^{\textcolor{Red}{\varheart}}$
}
{\large and\, \textcolor{Black}{C.~S.~Shahbazi}$^{\textcolor{CarnationPink}{\varheart}}$
}
 \vspace{0.5cm}

\renewcommand{\thefootnote}{\arabic{footnote}}
\vspace{0.1cm}

\small{$\textcolor{Red}{\hspace{-0.1cm} \varheart}$ \it Insituto de F\'isica Te\'orica UAM/CSIC \\ 
\hspace{-0.1cm}\it C/ Nicol\'as Cabrera, 13--15,  C.U.~Cantoblanco, 28049 Madrid, Spain}\\ \vspace{0.3cm}

\small{\hspace{-0.1cm} $\textcolor{CarnationPink}{ \varheart}$ \it Institut de Physique Theorique, CEA Saclay\\
\hspace{-0.1cm}\it CNRS URA 2306 , F-91191 Gif-sur-Yvette, France}
\vspace{1cm}
\hspace{-2cm}{\abstract{
{By choosing a particular, String Theory inspired, Special K\"ahler manifold, we are able to find a $\mathcal{N}=2$ four-dimensional ungauged Supergravity model that contains supersymmetric black hole solutions that violate the folk uniqueness theorems that are expected to hold in ungauged Supergravity. The black hole solutions are regular in the sense that they have a positive mass and a unique physical singularity hidden by an event horizon. In contradistinction to most examples already known in the literature, we find our solutions in a theory without scalar potential, gaugings or higher order curvature terms.}

}}
\end{center}

\setcounter{footnote}{0}

\pagestyle{plain}



\newpage
\section*{Introductory remarks}


The study of black holes in String Theory is an extremely active field of research\footnote{See \cite{Maldacena:1996ky,Skenderis:1999bs,Peet:2000hn, Mohaupt:2000mj,Das:2000su,Myers:2001ut, Shahbazi:2013ksa} and references therein.}. This is to be expected, since String Theory is thought to be a (the?) consistent theory of quantum gravity, and black holes seem to be the perfect theoretical laboratories in which to study the already old problem of the gravitational field quantization.

One way to study black holes in String Theory is by finding their \textit{classical} description as solutions to the different Supergravities that appear as the low-energy limit of String Theory compactifications. Over the last decades an enormous effort has been devoted to the study of the huge space of such String Theory-embeddable black hole solutions\footnote{See \cite{Andrianopoli:2006ub,Bellucci:2008jq,Dall'Agata:2011nh,Ortin:2010jm,Shahbazi:2013ksa} and references thereof.}. An effort which, after all, seems to be far from concluded.

Already from a purely classical point of view, black holes are extremely interesting objects whose physical properties have never ceased to amaze us since the very discovery of the Schwarzschild solution \cite{Schwarzschild:1916ae}. The existence of event horizons \cite{Hawking:1973uf}, the observation that certain black hole spacetimes contain completely causally disconnected regions \cite{Einstein:1935tc,Kruskal:1959vx}, or the discovery that black holes satisfy the laws of thermodynamics \cite{Bekenstein:1973ur,Hawking:1971tu,Bardeen:1973gs} are three good examples of such completely unexpected properties. 

Another interesting feature of these gravitational wonders comes from their exclusiveness. Indeed, it has been known for a long time now that all the stationary, asymptotically flat, black hole solutions to the Einstein-Maxwell theory, in a sort of general relativistic version of the Gauss law, are uniquely determined by a few parameters: their mass, their angular momentum and their electric and magnetic charges \cite{Jebsen:1921,Birkhoff,Israel:1967za,Carter:1971zc,Mazur:2000pn} \footnote{Thus, the only possible solution for a stationary, axisymmetric and electrovacuum black hole is given by the well-known Kerr-Newmann spacetime \cite{Mazur:1982db}.}.

The possible generalizations of these uniqueness (or \textit{No-Hair}) theorems to systems with more fields (such as scalars or non-Abelian vectors) has been an active area of research \cite{Bekenstein:1971hc,Coleman:1991ku,Bekenstein:1995un} since the proofs of the theorems for the simplest cases were carried out. On the other hand, the seek for counterexamples to the corresponding conjectured uniqueness theorems in such scenarios has also attracted a lot of attention, and produced some interesting results. In particular, it is now known that the No-Hair conjecture can be violated or circumvented in certain Einstein-Yang-Mills-Higgs systems (See \cite{Greene:1992fw,Kleihaus:2013tba,Anabalon:2012dw,Anabalon:2012ta,Anabalon:2013qua,Anabalon:2012sn,Winstanley:1995iq,Garcia-Compean:2013sla} and references therein) and in higher-curvature theories of gravity \cite{Mavromatos:1995fc,Kanti:1995vq}.

In this note we are going to construct a particular $\mathcal{N}=2$, $d=4$ ungauged Supergravity\footnote{Which in the case under consideration will be a particular instance of General Relativity coupled to an arbitrary number of Abelian vector fields and scalars.} model admitting pairs of supersymmetric, static, spherically symmetric and asymptotically flat black hole solutions sharing the same mass, charges and asymptotic values of the scalar fields, providing, to the best of our knowledge, the first counterexample to the corresponding uniqueness conjecture in the context of an ungauged Supergravity theory, and one of the first (some previous examples can be found in \cite{Anabalon:2013qua}) for a system without scalar potential, non-Abelian vector fields or higher-order curvature corrections.

In \cite{Bueno:2013psa} we obtained for the first time black hole solutions to a Type-IIA String Theory compactification on a Calabi-Yau manifold in the presence of non-perturbative corrections to the Special K\"ahler geometry of the vector multiplet sector \footnote{ The effective theory of Type-IIA String Theory on a Calabi-Yau manifold is four dimensional $\mathcal{N}=2$ ungauged Supergravity \cite{deWit:1983rz,Andrianopoli:1996cm}}. These black holes were given in terms of harmonic functions on euclidean $\mathbb{R}^{3}$, as it must be for supersymmetric black hole solutions of ungauged four dimensional Supergravity \cite{Huebscher:2006mr,Meessen:2010fh}, but they also contained a special function called the \textit{Lambert function}\footnote{See appendix \ref{sec:lambert}.}. As we argued in \cite{Bueno:2013psa}, the fact that the Lambert function is multivalued opened up the possibility of using its different branches to build inequivalent black hole solutions with the same conserved charges at infinity. However, such possibility was forbidden by the large volume compactification limit we assumed to hold through all the calculations: that limit only allowed us to consider solutions such that the argument of the Lambert function lied into a set of values for which the function was uniquely valued. 

Inspired by this result, we are going to construct a particular Supergravity model that can be analitically solved, and such that its supersymmetric black hole solutions share some of the characteristics of those in \cite{Bueno:2013psa}, but without any approximation involved. In particular, we will be able to construct solutions whose metric and scalars will depend on the Lambert function. In this case both branches will be available, and we will show how to construct a family of pairs of inequivalent solutions, providing a violation of the conjecture. 

In order to illustrate the result, we will show an explicit example for a model with two scalar fields. We will find that both solutions are regular, in the sense that the only physical singularity of the space-time will be hidden by an event horizon of non-zero positive area (for each solution in the pair). However, we will also see that the Special K\"ahler metric will not be positive definite (just like happens in other counterexamples to the conjecture \cite{Mavromatos:1995fc}) when evaluated on our solutions or, equivalently, that the energy-momentum tensor of at least one of the scalars will not satisfy in general the NEC. In this respect, and although the No-Hair conjecture does not make in principle reference to stability issues, it is fair to say that the spirit of the conjecture seems to remain partially alive.

The structure of the paper is the following. In section \ref{sec:generaltheory} we explain the structure of the bosonic sector of $\mathcal{N}=2, d=4$ Supergravity. In section \ref{sec:HFGK} we briefly review the H-FGK formalism, essential for the construction of the black hole solutions. In section \ref{sec:prep} we motivate the Supergravity model that we consider and we illustrate how we found it. In section \ref{sec:susysolution} we explicitly construct the supersymmetric solution without making any approximation. In sec \ref{sec:nohair} we explain how the family of pairs of solutions that we have constructed in section \ref{sec:susysolution} can be used to violate the No-Hair theorem, and finally, in section \ref{sec:examples}, we present an explicit pair of supersymmetric solutions, with the same mass, charges, and asymptotic value for the scalars at infinity.

\section{$\mathcal{N}=2, d=4$ Supergravity}
\label{sec:generaltheory}

\hspace{0.4cm}$\mathcal{N}=2,\, d=4$ Supergravity stands for any four-dimensional field theory invariant under the action of two independent local Supersymmetry generators \cite{Andrianopoli:1996cm}. Due to the $\mathbb{Z}_2$ symmetry $\phi_B\rightarrow \phi_B$ (bosonic fields), $\phi_F \rightarrow - \phi_F$ (fermionic fields) present in any Supergravity action, setting the fermions to zero is always a consistent truncation of the theory, which we will assume henceforth. We will restrict also to theories containing terms only up to two derivatives. The bosonic sector of any $\mathcal{N}=2, d=4$ Supergravity  can be written in that case as follows \cite{deWit:1983rz,Andrianopoli:1996cm}

\begin{align}
\label{eq:action}\notag
S= \int d^{4}x \sqrt{|g|}\,
\left\{
R
+h_{uv}(q)\partial_{\mu}q^u \partial^{\mu}q^{v}+\mathcal{G}_{i\bar{j}}(z,\bar{z})\partial_{\mu}z^{i}\partial^{\mu}\bar{z}^{\bar{j}}\right.\,\\  \left.+2\Im_{\Lambda\Sigma}(z,\bar{z})F^{\Lambda}{}_{\mu\nu}F^{\Sigma\, \mu\nu}-2\Re_{\Lambda\Sigma}(z,\bar{z})F^{\Lambda}{}_{\mu\nu} \star F^{\Sigma\, \mu\nu}
\right\}\, , \, \,\,
\end{align}

\noindent
where $R$ denotes the scalar curvature of the Levi-Civita connection associated to the space-time metric $g$; $q^u$ ($u=1,...,4n_h$) denotes the hyperscalars, which parametrize a $4n_h$-dimensional Quaternionic manifold $\mathcal{HM}$ with Riemannian metric $h_{uv}(q)$; $z^i$ ($i=1,...n_v$) denotes the $n_v$ complex scalar fields of the vector multiplets of the theory, which parametrize the $n_v$-dimensional base of a Special K\"ahler bundle $\mathcal{SV}=\mathcal{SM}\otimes \mathcal{L}$ with structural group Sp$(2n_v+2,\mathbb{R})\times$U$(1)$ and Riemannian metric $\mathcal{G}_{i\bar{j}}(z,\bar{z})$; $F^{\Lambda}_{\mu\nu}=2\partial_{[\mu}A^{\Lambda}_{\nu]}$ denote the field strengths of the $\Lambda=0,...,n_v$ 1-form connections $A^{i}$ of the vector multiplets plus the graviphoton $A^0$; and $\Im _{\Lambda\Sigma}\equiv \Im \frak{m} \mathcal{N}_{\Lambda\Sigma}(z,\bar{z})$, $\Re \equiv \Re \frak{e}_{\Lambda\Sigma} \mathcal{N}_{\Lambda\Sigma}(z,\bar{z})$ stand for the imaginary (negative definite) and real parts of the symplectic \textit{period} matrix $\mathcal{N}$, which determines the couplings of the 1-form connections $A^{\Lambda}$  to the scalars of the vector multiplets.

The hyperscalars $q^u$ are only coupled to themselves (and of curse to gravity) and, as a consequence, they can always be consistently fixed to constant values $q^u=q^u_0$. This simply means that the equations of motion for the hyperscalars  $q^{u}$ always admit the constant solution $q^u=q^u_0$. Of course, one may try to \textit{turn them on}, but it has been argued that no regular\footnote{That is, with the black hole singularity hidden by an event horizon.} black hole solutions with non-trivial hyperscalars can exist, since they would develop \textit{scalar hair}.

Supersymmetry constrains the couplings and kinetic terms of all the fields of the theory in a very particular way, which is beautifully codified in the language of Special Geometry\footnote{For an introduction to Special Geometry see, e.g. \cite{Andrianopoli:1996cm,Shahbazi:2013ksa}} for the vector multiplet sector. Indeed, the bosonic action in the absence of hyperscalars is determined as soon as we choose a holomorphic section $\Omega\in \Gamma(\mathcal{SV})$ or, equivalently when it exists, a homogeneous function $\mathcal{F}(\mathcal{X})$ of degree 2, called \textit{prepotential}, from which $\mathcal{G}_{i\bar{j}}$ and $\mathcal{N}_{\Lambda\Sigma}$ can be obtained as

\begin{eqnarray}
\mathcal{G}_{i\bar{j}}&=&-\partial_i\partial_{\bar{j}} \ln\left[i \left[\bar{\mathcal{X}}^{\Lambda}\partial_{\Lambda}\mathcal{F}-\mathcal{X}^{\Lambda}\partial_{_{\Lambda}}\bar{\mathcal{F}} \right] \right]\, ,\\
\mathcal{N}_{\Lambda\Sigma}&=&\partial_{\Lambda}\partial_{\Sigma}\bar{\mathcal{F}}+2i\frac{\Im \frak{m}(\partial_{\Lambda}\partial_{\Lambda^{\prime}}\bar{\mathcal{F}})\mathcal{X}^{{\Lambda}^{\prime}}\Im \frak{m}(\partial_{\Sigma}\partial_{\Sigma^{\prime}}\bar{\mathcal{F}})\mathcal{X}^{{\Sigma}^{\prime}}}{\mathcal{X}^{\Omega}\Im \frak{m}(\partial_{\Omega}\partial_{\Omega^{\prime}}\bar{\mathcal{F}})\mathcal{X}^{\Omega^{\prime}}}\,, \,\,\, \,\,\, \,\,\, \,\,
\end{eqnarray}

\noindent
where $\mathcal{X}^{\Lambda}$ are homogeneous coordinates on the scalar manifold, related to the $z^i$ by
\begin{equation}
z^i=\frac{\mathcal{X}^{i}}{\mathcal{X}^0}\, ,
\end{equation}
and where we have used the notation $\partial_{\Lambda}\equiv \frac{\partial}{\partial\mathcal{X}^{\Lambda}}$.

Therefore, it should be clear that if we choose a second degree homogeneous function $\mathcal{F}$, we automatically fix an $\mathcal{N}=2,\, d=4$ Supergravity theory coupled to vector multiplets. On the other hand, it is also reasonable to expect that not every election of prepotential will correspond to a Supergravity theory susceptible of being embedded in String Theory.

\section{H-FGK formalism}
\label{sec:HFGK}

\hspace{0.4cm}The most general static and spherically symmetric solution to (\ref{eq:action}) takes the form
\cite{Shahbazi:2013ksa,Ferrara:1997tw,Meessen:2011aa}
\begin{equation}
\label{eq:generalbhmetric}
\begin{array}{rcl}
\mathbf{g}
& = &
e^{2U(\tau)} dt\otimes dt - e^{-2U(\tau)} \gamma_{\underline{m} \underline{n}}
dx^{\underline{m}}\otimes dx^{\underline{n}}\, ,  \\
& & \\
\gamma_{\underline{m}\underline{n}}
dx^{\underline{m}} \otimes dx^{\underline{n}}
& = & \displaystyle
\frac{r_{0}^{2}}{\sinh^{2} r_{0}\tau}\left[ \frac{r_{0}^{2} }{ \sinh^{2}r_{0}\tau} d\tau\otimes d\tau
+
\displaystyle h_{S^2}\right]\, , \\
& & \\
h_{S^2}& = &d\theta \otimes d\theta +\sin^2\theta d\phi\otimes d\phi\, ,
\end{array}
\end{equation}

\noindent
where $\tau$ is the radial coordinate and $r_0$ is the \textit{non-extremality parameter} (which parametrizes \textit{how non-extremal} the black hole is) when (\ref{eq:generalbhmetric}) does in fact correspond to a black hole spacetime. In such a case, the exterior of the event horizon is covered by $\tau\in(-\infty,0)$, with the event horizon corresponding to $\tau\rightarrow -\infty$ and spacial infinity at $\tau\rightarrow 0^-$. The inner part of the Cauchy horizon is covered by $\tau\in(\tau_s,\infty)$, with the inner horizon at $\tau\rightarrow \infty$ and the singularity at $\tau=\tau_s$ for a certain positive and finite $\tau_s$ \cite{Galli:2011fq}.

We assume a static and spherically symmetric spacetime, as well as exclusively radial dependence for all the fields of the theory. In this case, the Maxwell equations can be integrated explicitly, and the vector fields can be written as functions of $\tau$ and the symplectic vector $\mathcal{Q}^M$ of electric $q_{\Lambda}$ and magnetic $p^{\Lambda}$ charges, $\mathcal{Q}^M\equiv\left(p^{\Lambda}, q_{\Lambda} \right)^T$. Indeed, let $\Psi \equiv (\psi^{\Lambda},\chi_{\Lambda})^{T}$ be a symplectic vector whose components are the time components (which are the only non-vanishing ones) of the electric $A^{\Lambda}$ and magnetic $A_{\Lambda}$ vector fields. Then, it can be shown that

\begin{equation}
\Psi^{M} = \int\frac{1}{2} e^{2U} \mathcal{M}^{MN}\mathcal{Q}_{N}d\tau\, ,
\end{equation}
\noindent
where $\mathcal{M}_{MN}$ is a symplectic and symmetric matrix constructed from the couplings of the scalars and the vector fields as
\begin{equation}
\label{M}
\left(\mathcal{M}_{MN}(\mathcal{N}) \right)
\equiv
\left(
\begin{array}{cc}
\left(\Im+\Re \Im^{-1}\Re\right)_{\Lambda\Sigma}  & -\left(\Re \Im^{-1}\right)_{\Lambda}^{\Sigma} \\
& \\
-\left(\Im ^{-1}\Re\right)_{\Sigma}^{\Lambda} & \left(\Im^{-1} \right)^{\Lambda\Sigma} \\
\end{array}
\right)\, .
\end{equation}
The (bosonic sector of the) four-dimensional $\mathcal{N}=2$ Supergravity action coupled to vector multiplets can be shown to be equivalent, assuming the space-time background given by (\ref{eq:generalbhmetric}), to the one-dimensional effective FGK action \cite{Ferrara:1997tw} for the $(2n_v+1)$ real fields $z^i(\tau)$ and $U(\tau)$

\begin{equation}
\label{FGK}
I_{\text{FGK}}[U,z^i]=\int d\tau \left\{ (\dot{U})^2+\mathcal{G}_{i\bar{j}}\dot{z}^i\dot{\bar{z}}^{\bar{j}}-e^{2U}V_{\rm bh}(z,\bar{z},\mathcal{Q}) \right\}\, ,
\end{equation}

\noindent
together with the \textit{Hamiltonian constraint}, which encondes the explicit independence of the effective lagrangian with respect to $\tau$

\begin{equation}
\label{haml}
(\dot{U})^2+\mathcal{G}_{i\bar{j}}\dot{z}^i\dot{\bar{z}}^{\bar{j}}+e^{2U}V_{\rm bh}(z,\bar{z},\mathcal{Q})=r_0^2\, .
\end{equation}

\noindent
In the previous expressions, $V_{\rm bh}$ is the so-called \textit{black hole potential}, which is defined by \cite{Ferrara:1997tw}

\begin{eqnarray}
\label{VBH}
V_{\rm bh}(z,\bar{z},\mathcal{Q})&\equiv& \frac{1}{2}\mathcal{M}_{MN}(\mathcal{N})\mathcal{Q}^M\mathcal{Q}^N\, .
\end{eqnarray}

\noindent
As we have said, the non-dependence of the effective FGK lagrangian on $\tau$ makes the corresponding Hamiltonian constant. In fact, the dimensional reduction over (\ref{eq:generalbhmetric}) imposes such constant to be precisely the square of the non-extremality parameter $r_0^2$.

The H-FGK formalism \cite{Galli:2011fq,Meessen:2011aa,Mohaupt:2011aa,Galli:2012ji,Bueno:2013pja} consists of a change of variables from $\left(U,z^{i}\right)$ to a new set of $(2n_v+2)$ variables $H^M(\tau)$ which transform under a symplectic, linear, representation of the U-duality group of the theory, and become harmonic functions in $\mathbb{R}^3$ in the supersymmetric case.  The equations of motion in the new variables $H^{M}(\tau)$ read
\begin{equation}
\label{eq:eoms}
g_{MN}\ddot{H}^{N}
+(\partial_{N}g_{PM}-\tfrac{1}{2}\partial_{M}g_{NP})\dot{H}^{N}\dot{H}^{P}
+\partial_{M}V=0\, ,
\end{equation}

\noindent
together with the \textit{Hamiltonian constraint}
\begin{equation}
\label{eq:hamiltonianconstraint}
\tfrac{1}{2}g_{MN}
\dot{H}^{M}\dot{H}^{N}
+V
+
r^{2}_{0}
=
0\, ,
\end{equation}

\noindent
where the (non invertible) metric $g_{MN}(H)$ and the potential $V(H)$ of the H-FGK effective action
are given in terms of the so-called \textit{Hesse potential} $\mathsf{W}(H)$ by
\begin{eqnarray}
g_{MN}(H) 
& \equiv &
\partial_{M}\partial_{N}\log{\mathsf{W}}
-2\frac{H_{M}H_{N}}{\mathsf{W}^{2}}\, ,
\\
& & \nonumber \\
\label{eq:potential}
V(H) & \equiv & 
\left\{
-\tfrac{1}{4}\partial_{M}\partial_{N}\log{\mathsf{W}}
+\frac{H_{M}H_{N}}{\mathsf{W}^{2}}
\right\}\mathcal{Q}^{M}\mathcal{Q}^{N}\, .
\end{eqnarray}
The relation between the Hesse potential, the $H^M$ variables and the covariantly-holomorphic symplectic section $\mathcal{V}^M$ is given by
\begin{equation}
\label{eq:W(H)}
\mathsf{W}(H) \equiv \tilde{H}_{M}(H)H^{M} = e^{-2U}, ~~~\tilde{H}^M+i H^M =\mathcal{V}^M/X\, ,
\end{equation}

\noindent
where $X$ is a complex variable with the same K\"ahler weight as $\mathcal{V}^M$, making the quotient $\mathcal{V}^M/X$ K\"ahler invariant. $\tilde{H}^M(H)$ stands for the real part $\left(\tilde{H}^M\right)$ of $\mathcal{V}^M/X$ written as a function of the imaginary part $H^M$, something that can always be done by solving the so-called \textit{stabilization equations}.

The effective theory is now expressed in terms of $2\left(n_{v}+1\right)$ variables $H^M$\footnote{Notice that the H-FGK formalism introduces an extra degree of freedom. As a consequence, the H-FGK action ejoys a gauge symmetry which, by gauge fixing, allows to get rid of it \cite{Galli:2012ji}.} and depends on $2\left(n_{v}+1\right)+1$ parameters: $2\left(n_{v}+1\right)$ charges $\mathcal{Q}^M=\left(p^{\Lambda},\, \, q_{\Lambda}\right)^T$ and the non-extremality parameter $r_0$, from which it is possible to reconstruct the solution in terms of the original fields of the theory (that is, the space-time metric, the scalars and the 1-form connections).

\section{A stringy motivation for the model}
\label{sec:prep}

The purpose of this letter is to study the supersymmetric black hole solutions of a particular $\mathcal{N}=2$ four dimensional ungauged Supergravity coupled to vector multiplets, which we will find to \textit{violate} the folk uniqueness theorems that are supposed to hold in unaguged four-dimensional Supergravity. Of course, such model did not appear out of the blue, but it has his seed and motivation in our previous paper \cite{Bueno:2013psa}.  In \cite{Bueno:2013psa} a new class of supersymmetric black hole solutions of type-IIA String Theory compactified to four dimensions on a Calabi-Yau manifold in the presence of non-perturbative stringy corrections was obtained. The supersymmetric solution was given by

\begin{equation}
\label{eq:phy}
e^{-2U}=\tilde{H}_iH^i \, ,~~z^i=i\frac{H^i}{\tilde{H}^0}\, ,
\end{equation}
\noindent
where
\begin{equation}
\label{roW}
\tilde{H}^0=\frac{\pi \hat{d}_lH^l}{ W_a\left(s_a \sqrt{\frac{3\hat{n}(\hat{d}_nH^n)^3}{2\kappa^0_{ijk}H^i H^j H^k}}\right)} \, ,
\end{equation}
\begin{equation}
\label{hi}
\tilde{H}_i=\frac{1}{2}\kappa^{0}_{ijk} \frac{H^j H^k}{\pi \hat{d}_lH^l}W_a\left(s_a \sqrt{\frac{3\hat{n}(\hat{d}_mH^m)^3}{2\kappa^0_{pqr}H^p H^q H^r}}\right)\, ,
\end{equation}
\noindent
and
\begin{equation}
\label{eq:universalsusy}
H^{i} = a^{i}-\frac{p^{i}}{\sqrt{2}}\tau\, ,
\end{equation}

\noindent
where $W_a(x) ~(a=0,-1)$ was any of the two real branches of \textit{Lambert's $W$ function}, and $s_a=\pm1$\footnote{See appendix \ref{sec:lambert}.}.

In order to solve the involved stabilization equations and obtain (\ref{roW}), (\ref{hi}) and (\ref{eq:universalsusy}) we were forced to consider the large volume limit $\Im \frak{m} z^i\rightarrow \infty$ of the compactification, where certain simplifications could be made. As a consequence, the approximation $\Im \frak{m} z^i\rightarrow \infty$ had to be also imposed on the solution. As explained in \cite{Bueno:2013psa}, only one of the two real branches of the $W$ function (the one with $a=0$) was consistent with such condition, which also implied the argument of $W_{0}(x(\tau))$ to be positive. We argued how, had not this condition been present, we could have tried to build two different solutions solving the same equations of motion, by choosing $W_{0}(x(\tau))$ or $W_{-1}(x(\tau))$ through (\ref{roW}), (\ref{hi}) and (\ref{eq:universalsusy}).

In fact, we could have assigned, through a suitable election of the parameters available in the solution, the near horizon ($\tau \rightarrow -\infty$) and asymptotic ($\tau \rightarrow 0$) limits of the argument $x(\tau)$ of $W_0(x(\tau))$ and $W_{-1}(x(\tau))$ to any pair of values chosen at will. In particular, we could have selected $x(\tau=0)=-1/e$ and $\lim_{\tau \rightarrow -\infty} x(\tau)=\beta$, $\beta \in (-1/e,0)$, and then the solution built with $W_{0}(x(\tau))$ and the one constructed with $W_{-1}(x(\tau))$ would have had exactly the same asymptotic behaviour, but different profiles away from infinity (note also (\ref{sec:lambert}) that $W_0$ and $W_{-1}$ are not even symmetric, in contradistinction to the branches of other real multivalued functions such as the inverse trigonometric functions). That is, we would have been dealing with two completely different regular solutions with the same mass $M$, charges and asymptotic values of the scalar fields, in contradiction with the aforementioned conjecture. Let us state that when we write \textit{regular}, we mean a black hole solution with positive mass $M$ such that there is a unique physical singularity in the space-time and it is hidden by an event horizon with non-zero, positive area.

In order to accomplish the construction of our solutions, we are going to somewhat forget about String Theory and propose a prepotential which we can solve exactly, and such that the corresponding supersymmetric solutions enjoy the same desirable properties as the String-Theory-forbidden ones of \cite{Bueno:2013psa}. In particular, we will use the same truncation in the $H$-variables, to wit

\begin{equation}
\label{eq:truncation}
H^0 = H_0 = H_i = 0,~~p^{0} = q_{0} = q_{i} = 0 \, .
\end{equation}

\noindent
In addition, we want the Lambert function to appear when solving the corresponding $0$-electric component of the stabilization equations. We have found that the following prepotential fulfils the required conditions

\begin{eqnarray}
\label{eq:IIaprepotentialHsi0}
F(\mathcal{X}) =\, n \left[ d_n \mathcal{X}^n\right] \left[ \mathcal{X}^0 e^{2i d_l \frac{\mathcal{X}^l}{\mathcal{X}^0}} - 2i \left[ d_m \mathcal{X}^m\right]Ei\left[2i d_l \frac{\mathcal{X}^l}{\mathcal{X}^0}\right]  \right]- d_{ijk} \frac{\mathcal{X}^i \mathcal{X}^j \mathcal{X}^k}{\mathcal{X}^0} \, ,
\end{eqnarray}

\noindent
where $Ei(z)$ is the \textit{exponential integral function}\footnote{See appendix \ref{expint}.}, and $d_{ijk}=d_{(ijk)}$, $n$ and $d_i$\footnote{The indices $i,j,k,l\dots$ run from 1 to a fixed arbitrary positive integer $n_v$.} are now arbitrary constants not constrained by any String Theory requirement, since we are considering a purely Supergravity model.

In the next section we are going to obtain the supersymmetric black hole solutions corresponding to the four dimensional $\mathcal{N}=2$ Supergravity theory defined by (\ref{eq:IIaprepotentialHsi0}), assuming the truncation (\ref{eq:truncation}).

\section{The supersymmetric solution}
\label{sec:susysolution}


In the H-FGK formalism, it is trivial to see that any $\mathcal{N}=2$, $d=4$ Supergravity model admits a solution for the $H^M$ variables given by

\begin{equation}
H^{M} = A^{M}-\frac{\mathcal{Q}^{M}}{\sqrt{2}}\tau\, ,
\end{equation}

\noindent
which turns out to correspond to a supersymmetric black hole \cite{Tod:1983pm,Gauntlett:2002nw,Meessen:2006tu,Huebscher:2006mr}. Using the truncation (\ref{eq:truncation}) we have

\begin{equation}
H^{i} = a^{i}-\frac{p^{i}}{\sqrt{2}}\tau\, ,\qquad H^{M} = 0\, ,\,\,\, M\neq i\, .
\end{equation}

\noindent
For the prepotential under consideration (\ref{eq:IIaprepotentialHsi0}), and the truncation (\ref{eq:truncation}), it is easy to see \cite{Bueno:2013psa} that the corresponding stabilization equation for $\tilde{H}^0$ is

\begin{equation}
\label{eq:F0}
\frac{\partial F(\mathcal{X})}{\partial \mathcal{X}^0}=d_{ijk} \frac{\mathcal{X}^i \mathcal{X}^j \mathcal{X}^k}{{\mathcal{X}^0}^2} +n \left[ d_n \mathcal{X}^n\right]  e^{2i d_l \frac{\mathcal{X}^l}{\mathcal{X}^0}}=0\, ,
\end{equation}

\noindent
which is solved by

\begin{equation}
\label{eq:R0}
\tilde{H}^0=\frac{d_lH^l}{ W_a\left(s_a \sqrt{\frac{ n(d_nH^n)^3}{d_{ijk}H^i H^j H^k}}\right)} \, .
\end{equation}

\noindent
This is precisely the same result that we found for the $\tilde{H}^0$ of the solution in the String Theory case \cite{Bueno:2013psa}, and which incorporates the Lambert function, as we wanted. The remaining stabilization equation to be solved is

\begin{eqnarray}
\label{eq:Fi}
\tilde{H}_{i} = \frac{\partial F(\mathcal{X})}{\partial \mathcal{X}^i}  \frac{e^{\frac{-\mathcal{K}}{2}}}{X}\, ,
\end{eqnarray}

\noindent
and its solution reads

\begin{eqnarray}
\label{eq:roiW}
\tilde{H}_i =  \frac{3d_{ijk}H^j H^k}{\tilde{H}^0} + n d_i \left[ e^{-\frac{2d_l H^l}{\tilde{H}^0}}\tilde{H}^0 + \left[4 d_mH^m\right] Ei\left[-\frac{2 d_q H^q}{\tilde{H}^0}\right]  \right] \, .
\end{eqnarray}

\noindent
$\tilde{H}_i$ becomes an explicit function of the $H^{i}$ once we substitute (\ref{eq:R0}) into (\ref{eq:roiW}). In any case the result is different from the corresponding one in the String Theory solution, which is to be expected since the model, although sharing some general characteristics, is different. The metric warp factor is hence given by

\begin{eqnarray}
\label{eq:roWw}
\displaystyle e^{-2U} =\,n \left[ d_nH^n\right] \left[ e^{-\frac{2 d_l H^l}{\tilde{H}^0}}\tilde{H}^0 +\left[4 d_mH^m\right] Ei\left[-\frac{2 d_q H^q}{\tilde{H}^0}\right]  \right]+ \frac{3d_{ijk}H^iH^j H^k}{\tilde{H}^0}\, ,
\end{eqnarray}

\noindent
whereas the scalars read

\begin{equation}
\label{eq:scalarsare}
z^{i}=\frac{\mathcal{X}^{i}}{\mathcal{X}^{0}} =i \frac{H^{i}}{d_lH^l}W_a\left(s_a \sqrt{\frac{ n(d_nH^n)^3}{d_{ijk}H^i H^j H^k}}\right)\, .
\end{equation}

\noindent
This completes the general construction of the supersymmetric solution. Of course, now we have to require, in order to have a regular solution, several conditions which will now be studied.


\subsection{Regularity conditions}


In order to have a regular solution the following requirements have to be satisfied:

\begin{enumerate}

\item The warp factor must be non zero, namely

\begin{equation}
e^{2U}>0\, ,\qquad\forall \,\, \tau\in (-\infty, 0)\, .
\end{equation}

\item The mass $M$ of the solution must be positive and finite

\begin{equation}
\label{eq:masscondition}
M\equiv \dot{U}(\tau \rightarrow 0)> 0\, .
\end{equation}
This requires a bit more explanation. Indeed, it turns out that the definition of the black hole mass involves derivatives of the Lambert function evaluated at $x(\tau=0)$, which will appear multiplicatively in the different factors of $\dot{U}(\tau)$. As we have sketched already and will explain in the next section, in order to jeopardize the No-Hair conjecture we want to fix the parameters of our solution in a way such that the argument of the Lambert function evaluated at spatial infinity ($\tau=0$) takes the value $-1/e$, where the two branches of $W$ make contact. However, it turns out that $W^{\prime}(x)$ diverges as $x\rightarrow -1/e$ (as explained in the appendix \ref{sec:lambert}). Fortunately, it is not difficult to cure this behaviour and get a positive (and finite) mass by choosing the parameters of the solution to be such that $\dot{x}(\tau)\overset{\tau \rightarrow 0 }{\longrightarrow} 0$ faster than $|W^{\prime}_{0,-1}(x)|\overset{x \rightarrow -1/e}{\longrightarrow}\pm \infty$. For instance, we can impose that the coefficient of order $\tau^0$ in the numerator of $\dot{x}(\tau)$ vanishes. As we will see in the explicit examples of section \ref{sec:examples}, this suffices to obtain a finite and positive mass for our pairs of inequivalent black holes.
\noindent

\item The K\"ahler potential must be consistently defined. That is

\begin{equation}
e^{\mathcal{-K}} = i\,\Omega_{M}\bar{\Omega}^M \, 
\end{equation}

\noindent
must be positive. For the prepotential (\ref{eq:IIaprepotentialHsi0}) the K\"ahler potential is given by

\begin{eqnarray}
\label{eq:eK}
 e^{-\mathcal{K}} &=& i d_{ijk} (z-\bar{z})^{i} (z-\bar{z})^{j} (z-\bar{z})^{k}  + i n d_{i}\left(z+\bar{z}\right)^{i} \left(e^{2id_{l} z^{l}} - e^{-2id_{l} \bar{z}^{l}}\right)\\ \notag &+& 4 n |d_{i} z^{i}|^2 \left( Ei \left[ 2 i d_{i} z^{i}\right] + Ei \left[ -2 i d_{i} \bar{z}^{i}\right] \right)\, .
\end{eqnarray}

\noindent
Since the supersymmetric solution that we have constructed has purely imaginary scalars, we can use $\bar{z}^{i} = -z^{i}$ to simplify this expression

\begin{equation}
\label{eq:eKreduced}
\frac{e^{-\mathcal{K}}}{8} = i d_{ijk} z^{i} z^{j} z^{k}+ n |d_{i} z^{i}|^2  Ei \left[ 2 i d_{i} z^{i}\right]\, .
\end{equation}

\end{enumerate}

\noindent
To summarize, if we obtain a solution such that the metric factor, the K\"ahler potential, and the mass are definite positive, we will have a regular black hole solution with a physical singularity hidden by an event horizon, and no other space-time singularities. 

\section{The violation of the No-Hair conjecture}
\label{sec:nohair}


The resolution of the stabilization equations given in section (\ref{sec:susysolution}) gives us the opportunity to build the supersymmetric solution either using $W_{0}$ (solution which we will denote by $S_{0}$) or $W_{-1}$ (solution which we will denote by $S_{-1}$). Therefore, in order to prepare the set up for the violation of the uniqueness conjecture, we need to construct a solution such that the argument of $W_a\left(s_{a} \sqrt{\frac{ n(d_nH^n)^3}{d_{ijk}H^i H^j H^k}}\right)$, which we denote by $x(\tau)$, lies entirely in the interval $(-1/e,0)$, only \textit{touching} the value $-1/e$ when $\tau = 0$, that is, at spatial infinity. Notice that if we want the argument $x(\tau)$ to be negative we have to chose $s_{0}=s_{-1}=-1$, which we will assume henceforth. This way, we will be able to construct two different black hole solutions that solve the same equations of motion, and have the same mass, charges and moduli at infinity, but however are different, since the profiles of $W_{0}$ and $W_{-1}$ are different (and asymmetric) when evaluated in $(-1/e,0)$. Hence, we need to impose

\begin{equation}
\label{eq:infcondition}
x(0) = -\sqrt{\frac{ n(d_n a^n)^3}{d_{ijk}a^i a^j a^k}} =-\frac{1}{e}\, ,
\end{equation}

\noindent
and

\begin{equation}
\label{eq:mcondition}
x(\tau) \in (\frac{-1}{e},0)\, ,\,\,\forall \,\, \tau\in (-\infty, 0)\, .
\end{equation}

\noindent
Of course, as explained in the previous section, in order to have a regular solution we need to impose $M>0$ and $e^{-2U}, e^{-\mathcal{K}}>0$ for $\tau\in (-\infty, 0)$. Assuming that (\ref{eq:masscondition}), (and the discussion under it) and (\ref{eq:infcondition}) hold, the value of the scalars at infinity as well as the mass, for both solutions $S_{0}$ and $S_{-1}$ will be given by

\begin{equation}
M =\dot{U}(\tau \rightarrow 0) \, ,\qquad \Im z_{\infty}^{i} =-\frac{a^{i}}{d_{l} a^{l}}\, .
\end{equation}

\noindent
In order to show that it is indeed possible (and actually easy) to choose the parameters available in the model in a way such that we can obey all the  conditions (regularity plus (\ref{eq:infcondition}) and (\ref{eq:mcondition})), in the next section we will explicitly construct a pair of solutions satisfying the required properties for a particular model with two scalar fields. 

Another issue, related to the stability of the solution, is the positive definiteness of the scalar metric $\mathcal{G}_{i\bar{j}}$ evaluated on the solution. Such a condition, which is related to the fulfilment of the NEC associated to the energy-momentum tensor of all the scalar fields in our solution, turns out to be difficult to satisfy. In particular, for the simple models in which we have worked out the explicit construction of pairs of solutions with the same masses, charges and asymptotic values of the moduli (like the one in section \ref{sec:examples}), the scalar metric turns out to have both positive and negative eigenvalues (for both solutions in each pair), meaning that some of the scalars in our solutions fail to satisfy the NEC (just like in other counterexamples to the No-Hair conjecture \cite{Mavromatos:1995fc}). At this point it is not clear to us whether this is a feature shared by all the possible solutions eluding the No-Hair conjecture susceptible of being constructed in our model (for any number of scalar fields), or not. This is an open question which could be addressed from different approaches \cite{lapolla}. On the one hand, one could always try to map (brute force-wise) the parameter space for models with different numbers of scalars, looking for a solution satisfying all the requirements but with a positive definite scalar metric. It would also be possible to consider other prepotentials giving rise to stabilization equations whose solutions involve multivalued functions, and study the situation therein. On the other hand, it might just be that our procedure of \textit{placing the spatial infinity} at the branch point of the Lambert function necessarily implies some unstable behaviour for the corresponding solutions, not incompatible with their regularity. This could be related to the structure of the attractor flows associated to each pair of solutions. Let us see how this works.


\subsection{Attractors}


Although both solutions $S_{0}$ and $S_{-1 }$ have exactly the same asymptotic limit $\tau\to 0$, since the flow is different, one should expect that the corresponding attractors $z_{0}$ and $z_{-1}$ are different. This is indeed the case; they are given by

\begin{equation}
z_{a}^{i} =  \frac{p^{i}}{d_l p^l}W_a\left(- \sqrt{\frac{ n(d_l p^l)^3}{d_{ijk}p^i p^j p^k}}\right)\, .
\end{equation}

\noindent
 This can be understood in the context of the basins of attractions \cite{Kallosh:1999mz}. Let us suppose that we impose

 \begin{equation}
x(0) = \alpha\, , \qquad \alpha\in (-\frac{1}{e}, \beta)\, ,
\end{equation}

\noindent
instead of $x(0)=-\frac{1}{e}$. Then $S_{0}$ and $S_{-1}$ have different asymptotic limits at spatial infinity. In particular, the asymptotic value of the scalars at infinity is different for  $S_{0}$ and $S_{-1}$. Therefore, we have two basins of attraction $B_{0}$ and $B_{1}$ such that the solution $S_{0}$ corresponds to $B_{0}$ and $S_{-1}$ corresponds to $B_{-1}$. What happens when we impose

 \begin{equation}
x(0) =-\frac{1}{e}\, 
\end{equation}

\noindent
is that we precisely choose a point which lies in $B_{0}$ and $B_{-1}$, that is, we choose a point in the common border of the two basins of attraction. As a result, we end up with two different solutions, with different attractors, which however have the same mass, charges and asymptotic values of the scalares at infinity. 

This standpoint suggests that there could be, in fact, some instability associated to our election of the Lambert's function argument at the branch point. If this were the case, it would simply mean that, just like appears to happen in other counterexamples available in the literature (but those usually in theories with scalar potential, gaugings or higher order curvature corrections), the No-Hair Conjecture remains robust when stability issues are considered.


\section{An explicit example}
\label{sec:examples}
Let us consider a model with two scalar fields $z^1$ and $z^2$. The warp factor of the spacetime metric and the scalars can be read off directly from $(\ref{eq:roWw})$ and $(\ref{eq:scalarsare})$ with
$H^{1} = a^{1}-\frac{p^{1}}{\sqrt{2}}\tau$, $H^{2} = a^{2}-\frac{p^{2}}{\sqrt{2}}\tau$. Imposing the regularity conditions, the correct asymptotic behaviour of the metric ($e^{2U}\overset{\tau \rightarrow 0}{\Longrightarrow} 1$) and choosing the parameters in the argument of the two branches of the Lambert function in the way explained in the previous section (and such that (\ref{eq:mcondition}) and (\ref{eq:infcondition}) hold), it is not difficult to construct solutions with the required properties (and which, in all the examples constructed automatically satisfy the condition $e^{-\mathcal{K}}>0$ $\forall \tau  \in (-\infty,0)$). Let us choose a particular model with $d_1=d_2=1$, $d_{122}=0$, $d_{222}\simeq -0,270$, $d_{211}\simeq 0,320$, $d_{111}\simeq -2,040$, $n\simeq -0,011$, and with the following constants for our solutions: $\Im \frak{m} z_{\infty}^1=-1/3$, $\Im \frak{m} z_{\infty}^2=-2/3$, $p^1=p^2=1 $. The explicit dependence on $\tau$ of the warp factor and the imaginary parts of our scalars for the examples at hand is in general very messy, so instead of reproducing it here, let us have a look at the corresponding plots for this particular example, for which the mass turns out to be $M=2/3$\\

\begin{figure}[h!]

  \centering
    \includegraphics[width=0.8\textwidth]{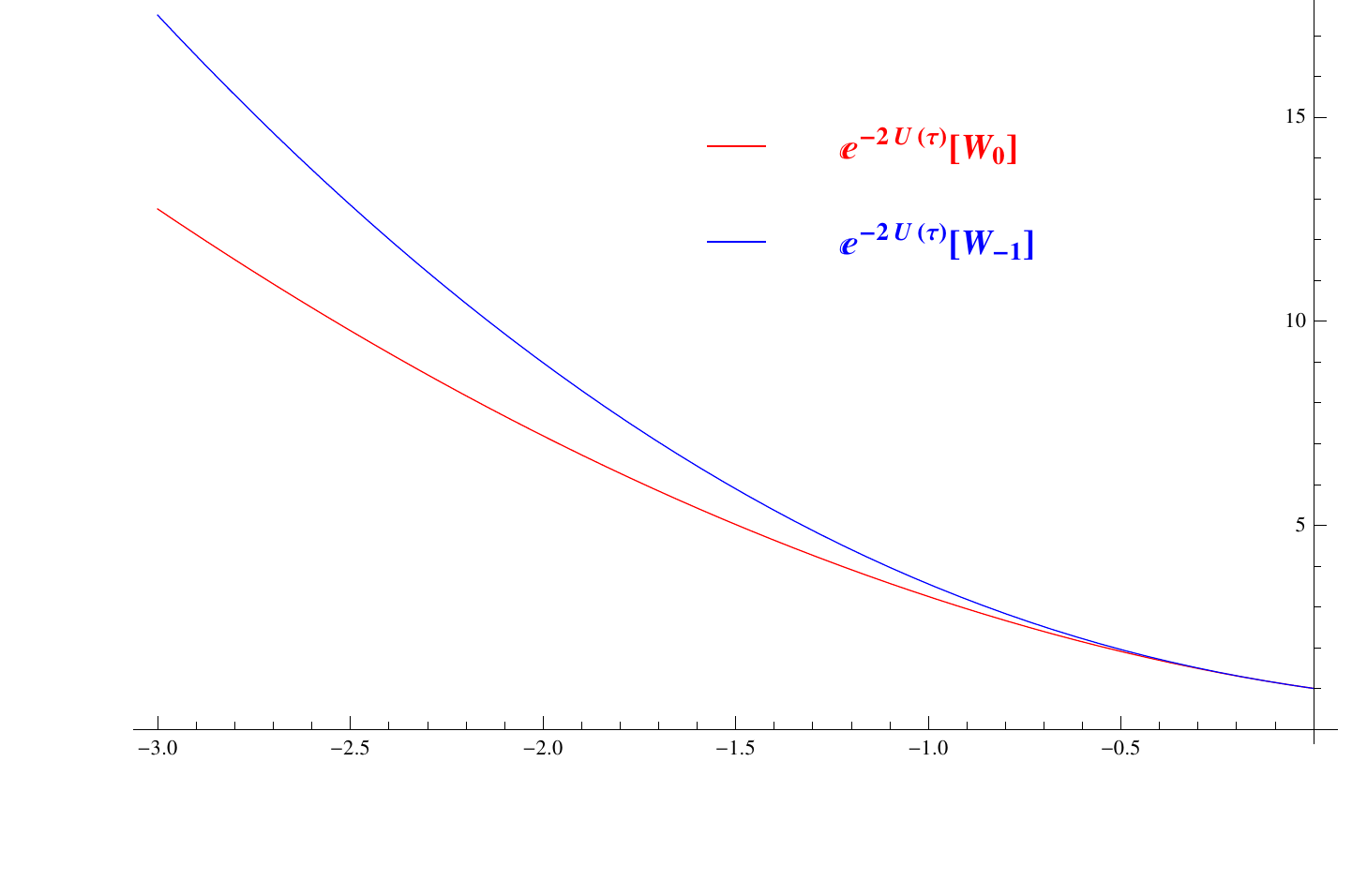}
    \vspace{-0.5cm}\caption{The profiles of the metric warp factors corresponding to the two solutions outside the event horizon $\tau \in(-\infty, 0)$. Both metrics asymptote to Minkowski spacetime at spatial infinity.}
\end{figure}

\newpage
\begin{figure}[h!]

  \centering
    \includegraphics[width=1\textwidth]{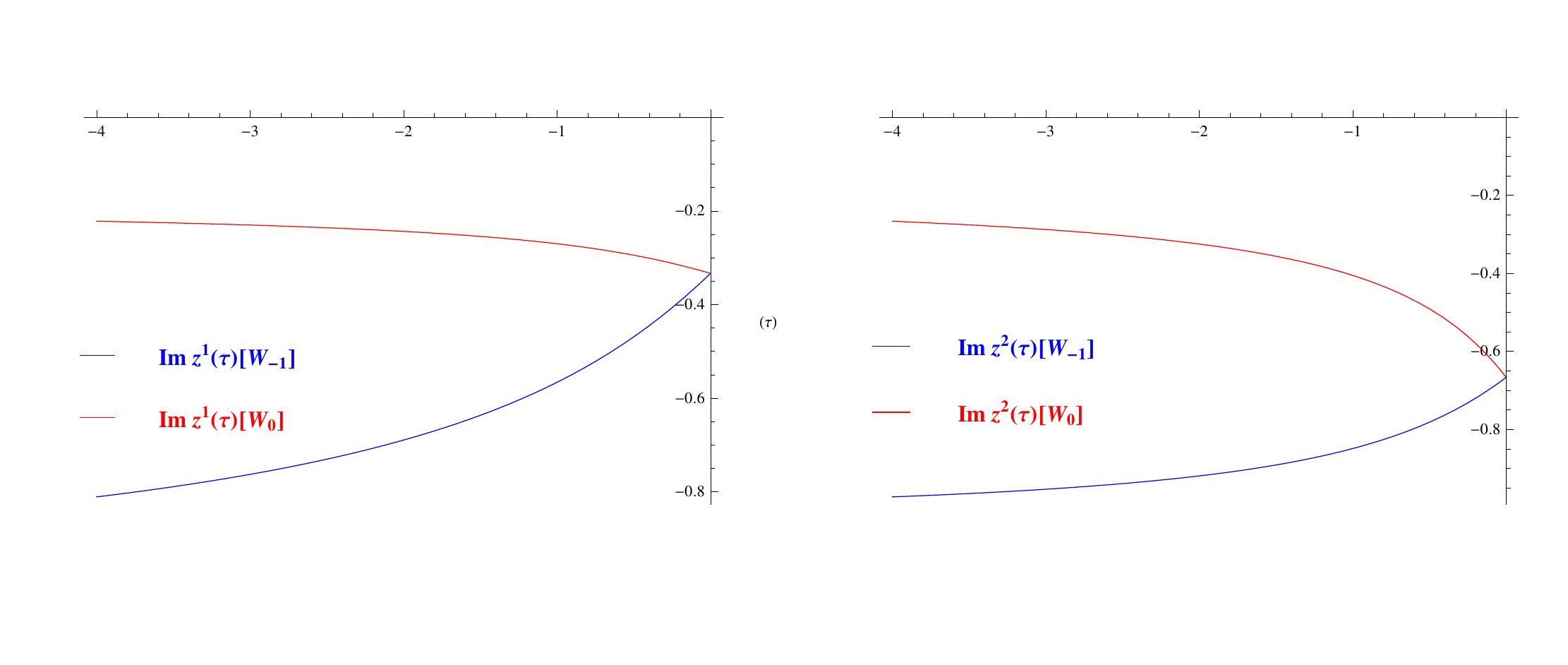}
    \caption{The profiles of the imaginary parts of the scalar fields outside the event horizon $\tau \in(-\infty, 0)$. As we can see, their asymptotic values $\Im \frak{m} z_{\infty}^1$ and $\Im \frak{m} z_{\infty}^2$ coincide for both solutions (recall that spatial infinity is at $\tau\rightarrow 0^-$).}
\end{figure}

As we can see, both solutions are completely regular, and share the same mass, charges, and asymptotic values of the scalars.



\section*{Acknowledgments}


We wish to thank Tom\'as Ort\'in for long and illuminating discussions, comments and for suggesting the interpretation of the solutions in the context of the basins of attraction. We are also thankful to Nick E. Mavromatos for useful comments related to the stability issue. PB whishes to thank the CERN Theory Division for its hospitality. This work has been supported in part by the Spanish Ministry of Science and Education grant FPA2012-35043-C02-01, the Comunidad de Madrid grant HEPHACOS S2009ESP-1473, and the Spanish Consolider-Ingenio 2010 program CPAN CSD2007-00042. The work of PB has been supported by the JAE-predoc grant JAEPre 2011 00452. The work of CSS has been supported by the JAEPre 2010 00613 grant and the ERC Starting Grant 259133. The authors acknowledge the support of the Spanish MINECO's \textit{Centro de Excelencia Severo Ochoa} Programme under grant SEV-2012-0249.

\appendix


\section{The Lambert W function}
\label{sec:lambert}


The \textit{Lambert W function} $W(z)$ was firstly introduced by Johann Heinrich Lambert in 1758 \cite{Lambert}. Along its history, it has found numerous applications in different areas of physics (mostly during the 20th century)\cite{Valluri,JM,Gardi:1998qr,Magradze:1998ng,Nesterenko:2003xb,Cvetic:2011vy,Sonoda:2013kia,Ashoorioon:2004vm,Mann,Mann:1996cb,Belyakova:2010iw}.

$W(z)$ is defined (implicitly) through the equation

\begin{equation}
\label{Wf}
z=W(z)e^{W(z)}\, ,~~ \forall z\in \mathbb{C}\,.
\end{equation}

\noindent
Since $f(z)=ze^{z}$ is not injective, $W(z)$ is not uniquely defined, and $W(z)$ stands for the whole set of branches solving (\ref{Wf}). For $W:\mathbb{R}\rightarrow \mathbb{R}$, $W(x)$ has two branches $W_0(x)$ and $W_{-1}(x)$ defined respectively in the intervals $x\in [-1/e,+\infty)$ and $x\in [-1/e,0)$ (See Figure 3). Both functions coincide in the branching point $x=-1/e$, where $W_0(-1/e)=W_{-1}(-1/e)=-1$. Therefore, the defining equation $x=W(x)e^{W(x)}$ admits two different solutions in the interval $x\in [-1/e,0)$.\\
\begin{figure}[h]
 \label{fig:u}
  \centering
    \includegraphics[scale=0.7]{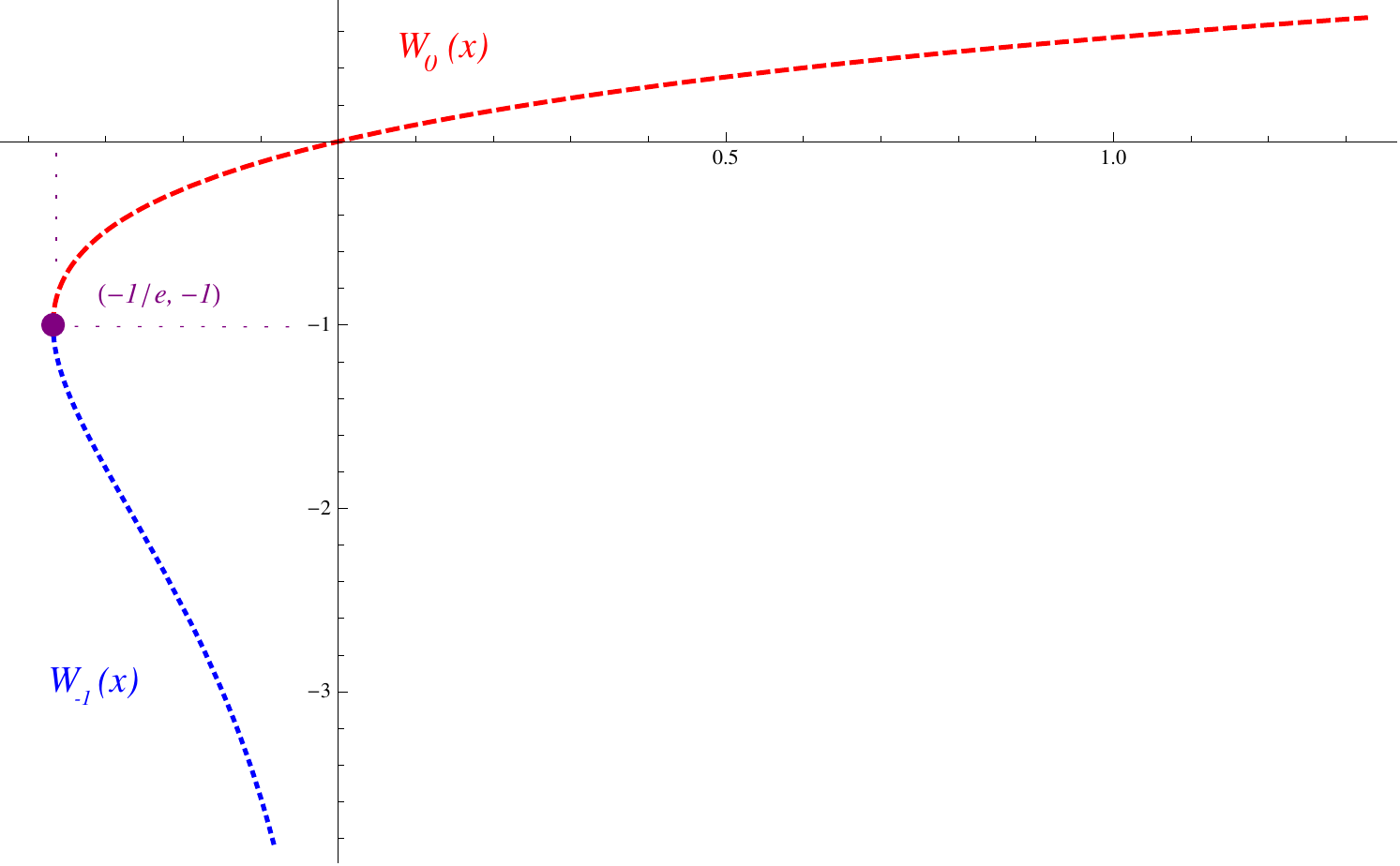}
 \caption{\small{The two real branches of $W(x)$.}}
\end{figure}

\noindent
The derivative of $W(z)$ reads

\begin{equation}
\frac{dW(z)}{dz}=\frac{W(z)}{z(1+W(z))},~~\forall z \notin \left\{0,-1/e \right\}; ~~ \frac{dW(z)}{dz}\bigg|_{z=0}=1 \, ,
\end{equation}

\noindent
and is not defined for $z=-1/e$ (the function is not differentiable there). At that point one finds

\begin{equation}
\lim_{x\rightarrow -1/e}\frac{dW_0(x)}{dx}=\infty,~~\lim_{x\rightarrow -1/e}\frac{dW_{-1}(x)}{dx}=-\infty \, .
\end{equation}


\section{The Exponential Integral function}
\label{expint}


The Exponential Integral $Ei\left[z\right]\, ,\,\, z\in\mathbb{C}$ is a special function on the complex plane. For real non-zero values $x$ it is defined as follows

\begin{equation}
Ei(x) = -\int^{\infty}_{-x}\frac{e^{-t}}{t} dt\, .
\end{equation}

\noindent
We only need the Exponential Integral function evaluated in the real numbers since in our solutions it appears only with a real argument, although in the definition of the prepotential (\ref{eq:IIaprepotentialHsi0}) it appears with an argument that can be in general complex.

\begin{figure}[h]
 \label{fig:Ei}
  \centering
    \includegraphics[scale=0.7]{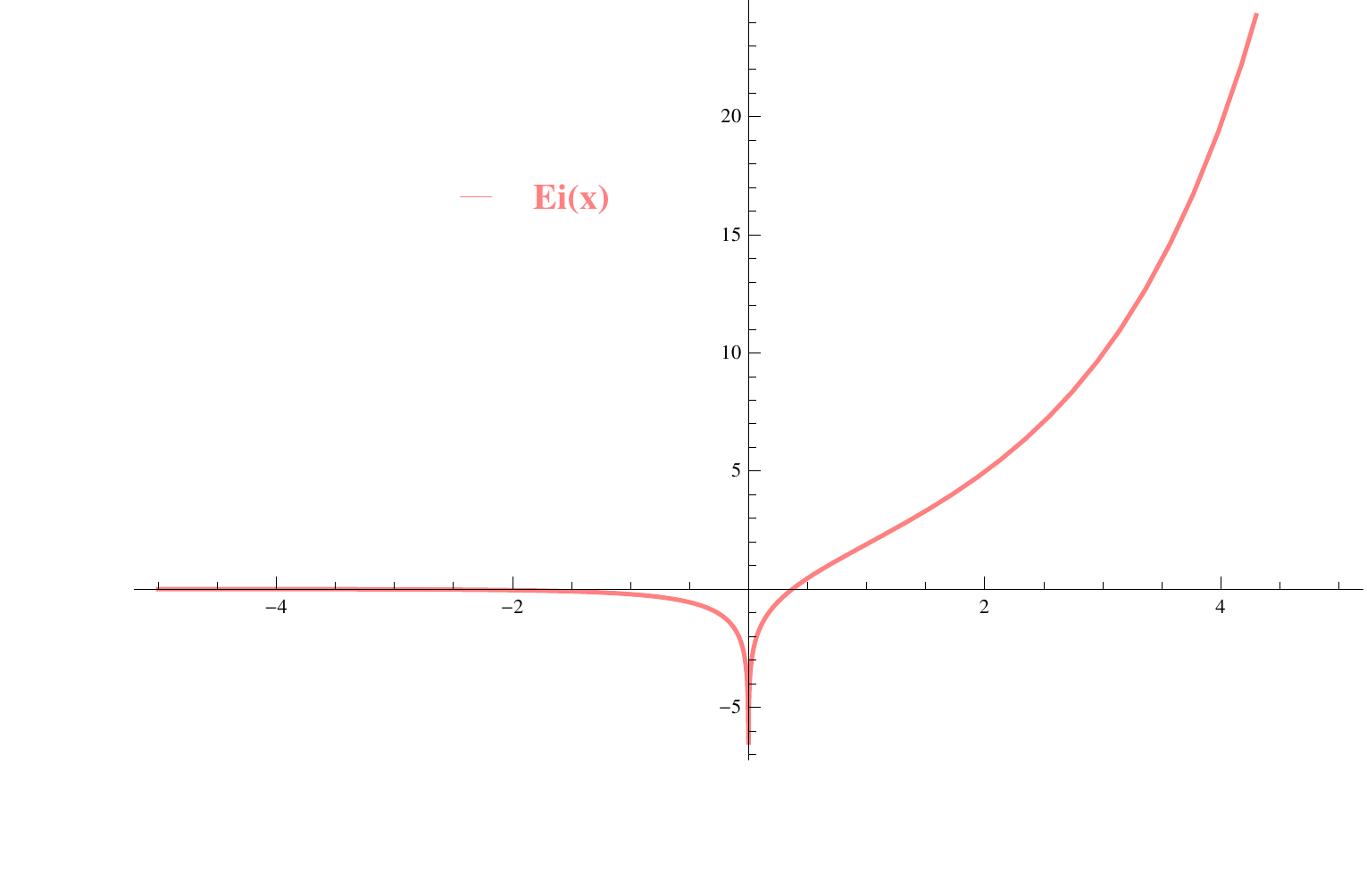}
 \caption{\small{The Exponential Integral function on the real axis.}}
\end{figure}

\noindent
$Ei(x)$ is negative for $x\in(-\infty,c)$, where $c\sim 0,375$, zero in $x = c$ and positive for $x>c$. In addition,  $\lim_{x\to 0}Ei(x) = -\infty$.

\renewcommand{\leftmark}{\MakeUppercase{Bibliography}}
\phantomsection
\addcontentsline{toc}{chapter}{References}
\bibliographystyle{JHEP}
\bibliography{References.bib}
\label{biblio}

\line(1,0){100}

\noindent
\textit{E-mail}:\\
 $\textcolor{Red}{\varheart}$\texttt{pab.bueno[at]estudiante.uam.es\\ $\textcolor{CarnationPink}{\varheart}$carlos.shabazi[at]cea.fr}\\
\noindent
\end{document}